\documentclass[onecolumn]{IEEEtran}
\usepackage{mathrsfs}
\usepackage{mathrsfs}
\usepackage{mathrsfs}
\usepackage{amssymb}
\usepackage{amsthm}
\usepackage{amsmath}
\usepackage{amsfonts}
\usepackage{mathrsfs}
\usepackage{amssymb}
\usepackage[dvips]{graphicx}
\usepackage{subfigure}
\usepackage{bm}
\usepackage{caption}
\usepackage{stfloats}
\usepackage{epsfig}
\usepackage{algorithm}
\usepackage{algorithmic}
\usepackage[dvips]{color}
\usepackage{subeqnarray}
\usepackage{cases}
\usepackage{cite}

\begin{document}

\title{Performance Analysis and Optimization for the MAC Protocol in UAV-based IoT Network}

\author{Bin Li, Xianzhen Guo, Ruonan Zhang, Xiaojiang Du, {\it Fellow, IEEE}, Mohsen Guizani, {\it Fellow, IEEE}}

\maketitle

\begin{abstract}

Unmanned aerial vehicles (UAVs) have played an important role in air-ground integration network. Especially in Internet of Things (IoT) services, UAV equipped with communication equipments is widely adopted as a mobile base station (BS) for data collection from IoT devices on the ground.
In this paper, we consider an air-ground network in which the UAV flies straightly to collect information from the IoT devices in a 2-D plane based on the CSMA/CA protocol. Due to UAV's continuous mobility, the communication durations of devices in different locations with UAV are not only time-limited, but also vary from each other. To analyze the throughput performance of uplink multiple access control (MAC) protocol, we propose a new analysis model to deal with the communications heterogeneity in the network. Firstly, we divide the devices in the coverage into different clusters according to their communication durations. Then, a quitting probability indicating the probability that a device quits the UAV's coverage at each time slot is clarified. A modified three-dimensional Markov chain model adopting the quitting probability and cluster division is developed for the performance analysis. Besides, we also propose a modified CSMA/CA protocol which fully considers the heterogeneity of the access time and adaptively allocates the time resource among the devices in different clusters. Finally, the effects of retry limit, initial contention window size, the density of the devices, UAV's speed and coverage area are discussed in the simulation section.

\end{abstract}


\IEEEpeerreviewmaketitle

\begin{figure}[b]
\vspace{-0mm}
\footnotesize{Copyright (c) 2015 IEEE. Personal use of this material is permitted. However, permission to use this material for any other purposes must be obtained from the IEEE by sending a request to pubs-permissions@ieee.org.

This work is partially supported by National Natural Science Foundation of China (Nos. 61601365, 61571370), Natural Science Basic Research Plan in Shaanxi Province (No. 2019JM-345), National Civil Aircraft Major Project of China (No. MIZ-2015-F-009), China Postdoctoral Science Foundation(No. 2018M641020), Science and Technology Research Program of Shaanxi Province (No. 2019ZDLGY07-10), Advance Research Program on Common Information System Technologies (No. 315075702).

Bin Li, Xianzhen Guo and Ruonan Zhang are with the Department of Communication Engineering, Northwestern Polytechnical University, Xi'an, China 710072 (e-mail: {\tt libin@nwpu.edu.cn}; {\tt xz.g@mail.nwpu.edu.cn}; {\tt rzhang@nwpu.edu.cn}).

Xiaojiang Du is with the Department of Computer and Information Sciences, Temple University, Philadelphia,
PA, USA. (e-mail: {\tt dxj@ieee.org}).

Mohsen Guizani is with Department of Computer Science and Engineering, Qatar University, Qatar, (e-mail: {\tt mguizani@ieee.org})

}
\end{figure}

\IEEEpeerreviewmaketitle

\section{Introduction}

Wireless communication with unmanned aerial vehicles (UAVs) has been a popular technology in air-ground integration network for military, public, and emergency applications \cite{b1}. Compared with the terrestrial communication system, UAV-based air-ground network is not only easy for deployment, but also has better wireless channels \cite{b2}, \cite{guo1}. Therefore, numerous UAV-related wireless communication systems have been developed \cite{d1,d2,d3,du1}. One typical application is the UAV-based air-ground IoT system \cite{du2}, in which UAV is deployed as a mobile BS flying over the IoT devices to collect the sensor information \cite{d4}, \cite{ad3}. In this way, the UAV-based network gets rid of the complex routing design and greatly improves the efficiency of information collection. However, there are still some challenges in analyzing and designing the UAV-based data collection system.

One of the challenges is that it is difficult to model the UAV-based communication system compared with the traditional static network due to UAV's mobility. Generally, the fixed-wing UAV is usually adopted in data collection scenario for its long endurance. This characteristic requires the UAV to fly continuously without hovering, causing the time-varying channels between the transceivers. In this case, the devices can only access the UAV when the UAV is close to them while the wireless links break if the UAV flies away, resulting in the limited communication time.
On the other hand, the UAV is usually equipped with directional antennas for energy-efficient transmission, forming a circular signal coverage on the ground. In this way, devices located in different areas can communicate with the UAV for different time, leading to the communication heterogeneity in the coverage. This heterogeneity causes an unfair accessing opportunity in the network. Specifically, the devices with longer communication time have more opportunities to transmit information while others may not be able to access the channel for data uploading due to the limited time. Therefore, how to select or design an appropriate multiple access control (MAC) protocol for uplink channel to tackle those issues is a challenging problem.

In fact, the Time Division ultiple Access (TDMA) scheme may not be an effective solution to deal with the heterogeneity of the access time. While the Frequency Division Multiple Access (FDMA) scheme is also not appropriate because it is difficult to maintain the stringent synchronization of the frequencies distributed to all the devices \cite{b4}. As a widely adopted MAC protocol, CSMA/CA could be a good choice in UAV-based data collection network for its flexible access strategy\cite{b5}. Multiple theoretical models for the performance analysis of CSMA/CA protocol in conventional communication system with static BS have been proposed \cite{b6}, \cite{b7}. However, as far as we know, there is no such a theoretical model to evaluate the performance of CSMA/CA in UAV-based data collection system.

Motivated by the challenges above, in this paper, we develop a theoretical model of CSMA/CA protocol in UAV-based system to analyze the network performance. In detail, we first divide the devices in the coverage into different clusters according to the communication durations. Then, we introduce a quitting probability indicating the probability that a device quits the UAV’s coverage at each time slot. Based on the quitting probability and cluster division, a modified three-dimensional Markov chain model is developed for performance analysis. According to the theoretical analysis, we also propose a new MAC protocol that fully considers the heterogeneity of the access time and adaptively allocates the time resource among the devices within difference clusters.

Our main novelty and contributions are summarized as follows.

\begin{itemize}
	\item We analyze the characters of the UAV-based IoT network and propose a new quitting probability and cluster division strategy to describe the devices access process by considering the location difference among the devices.
	
	\item Based on the quitting probability and cluster division, a modified three-dimensional Markov chain model is set up to calculate the saturation throughput of CSMA/CA protocol in UAV-based IoT network.
	
	\item Based on the quantitative analysis, we propose a modified CSMA/CA protocol, which fully considers the heterogeneity of the access time, to adaptively allocate the time resource among the devices in different clusters.
\end{itemize}

The rest of this paper is organized as follows. The related work is shown in Section~\ref{section:1}. System model is presented in Section~\ref{section:2}. Section~\ref{section:3} introduces the proposed 3-D Markov chain model. Section~\ref{section:4} analyzes the Markov model process and gives the computation of throughput. The modified CSMA/CA protocol is proposed in Section~\ref{section:5}. Numerical results are shown in Section~\ref{section:6}. Finally, Section~\ref{section:7} gives some conclusions about this paper.

\section{Related Work}\label{section:1}

In this section, we present the related work on UAV-based communications network.

\cite{b8} gave an extensive literature review about UAV communications network. It presented various communication schemes by considering the mobility of the BS. \cite{b9} proposed a basic framework for aerial data collection by considering the deployment of networks, node positioning, anchor points searching, fast path planning for UAV, and data collection schemes.
\cite{b111} proposed a framework in linear wireless sensor network based on UAV and developed three different network models based on UAV's movement strategy. In \cite{b10}, the general fading channel model for sensor nodes and links was considered. The sensors' wake-up scheme and UAV's trajectory were jointly optimized using a convex sub-optimal optimization solution for energy-efficiency. The UAV in \cite{b118} was assumed to be able to either cruise or hover while collecting the sensors' data. It focused on the UAV's flight time minimization problem for data collection over wireless sensor network. \cite{b113} analyzed the effects of some UAV mobility patterns such as angular mobility, square mobility, and circular mobility patterns on data collection in wireless sensor network. \cite{b117} studied the fairness-aware UAV-assisted data collection problem and proposed four data collection algorithms by considering the multi-data-rate transmission and the contact duration time between the sensors and the UAV. In \cite{ad4}, UAV was deployed as a mobile relay and the UAV's trajectory and energy and service time allocations were optimized jointly to minimize the average peak age-of-information for a source-destination pair.
\cite{ad1} proposed a AC-POCA algorithm based on partially overlapping channels and game theory to overcome the problem caused by the dynamic topology and high mobility nodes in a combined UAV and D2D based network. 

Aforementioned work focused on the system architecture or the design of UAV's trajectory. Few of them have analyzed the performance of the data collection protocol, which is important for guidance of UAV communication network design.

For MAC protocol performance ananlysis, \cite{b6} gave a simple but extremely accurate model to calculate the throughput of CSMA/CA protocol using Markov chain model. This model could be applied to the basic access, the RTS/CTS access mechanisms and the combination of these two schemes. \cite{b7} extended the multi-dimensional Markov model by considering the impacts of both non-ideal channel and capture effects in Rayleigh fading environment. It also proposed a new idle state to describe the unsaturated traffic condition. Both the saturated and unsaturated throughput could be calculated according to the analytical model. \cite{b11} took the effects of packet retry limits and transmission errors into account in order to obtain a more accurate model. Unfortunately, all the analytical models mentioned above are based on a static communication system.

For the UAV-based MAC protocols, \cite{bov} did a survey of MAC protocol for UAV-aided wireless sensor networks. Various MAC protocols were extensively investigated and compared with each other in terms of major features, pros, and cons. \cite{b12} adopted a conventional CSMA/CA and used a priority-based optimized frame selection scheme to propose an effective data gathering protocol. It divided the UAV's coverage into multiple frames which were assigned with different transmission priorities respectively. \cite{b13} proposed a novel data acquisition framework to increase the efficiency of the data gathering. In this framework, a priority-based frame selection was accepted. An adaptive hybrid beacon based MAC protocol was proposed in \cite{b5} by combining CSMA/CA with physical parameters based scheduling. \cite{b14} introduced a new MAC protocol based on IEEE 802.11 standard for data gathering in wireless sensor networks. This protocol gave a fair chance to each sensor to communicate with the drone and its performance was analyzed using NS-2 simulations.

However, the performance of the protocols proposed
in \cite{b5}, \cite{b12,b13,b14} was evaluated using simulation tools,
i.e. MATLAB or NS2. None of them gave a theoretical
model of the protocols’ throughput. In addition, only
the RTS/CTS mechanism was treated in those papers,
ignoring the basic access mechanism. 

\cite{b114} presented a collision-free MAC protocol called CF-MAC for UAVs Ad-Hoc networks. This protocol enabled the UAV to access the networks rapidly and the collision probability could be reduced to near zero by utilizing a region marking scheme. \cite{b1114} considered the delay tolerant transmission in UAV data collection networks. It proposed a delay tolerant multiple access control protocol called UD-MAC and different access priorities were set in UD-MAC to achieve superior access efficiency. In addition, some MAC protocols for UAV communication have been proposed by considering the effects of directional antennas. \cite{b112} proposed a position-prediction-based directional MAC protocol and a propagation delay-aware MAC scheme for long-distance UAV networks called LDMAC was given in \cite{b119}. \cite{b1110} presented a location orientated directional MAC protocol which incorporated the utilization of directional antennas and location estimation of the neighboring nodes. The MAC protocols in \cite{b114,b1114,b112,b119,b1110} were designed
for communications among UAVs in UAV ad-hoc network
instead of the UAV-based data collection scenario where the UAV
communicated with multiple IoT devices on the ground.

In our paper, we provide an theoretical model to compute the CSMA/CA protocol throughput in UAV data collection system. The proposed analysis model can be applied both to the RTS/CTS mechanism and the basic access mechanism. Moreover, considering the varied communication durations among devices, we propose a modified MAC protocol based on CSMA/CA protocol. According to the simulation results, the modified protocol can improve the throughput of the data collection system enormously.

\section{System Model}\label{section:2}

In this section, a scenario where a UAV collects data from the IoT devices on the ground is described. Then, a classic Markov chain model in CSMA/CA protocol is introduced briefly. After that, we design a strategy of the cluster division. Finally, a new concept called quitting probability in MAC protocol is defined and derived.

\subsection{UAV-based Air-ground Network Model}
We consider a scenario where a UAV equipped with a data collector tries to collect the data from a set of IoT devices distributed normally on a 2-D plane. Assume the UAV's position is $({x_u},{y_u},{z_u})$. Let $({x_i},{y_i})$ denote the location of device $s_i, i=1,2,...,K$, where $K$ is the total number of IoT devices in the UAV's coverage.
For a certain period of time, the UAV is assumed to move in a straight trajectory over the devices because the UAV is not supposed to turn too frequently for the aerodynamic limitation. All the devices in the UAV's coverage try to access the UAV randomly.

\begin{figure}[!htp]
	\centering
	\includegraphics [width=9cm]{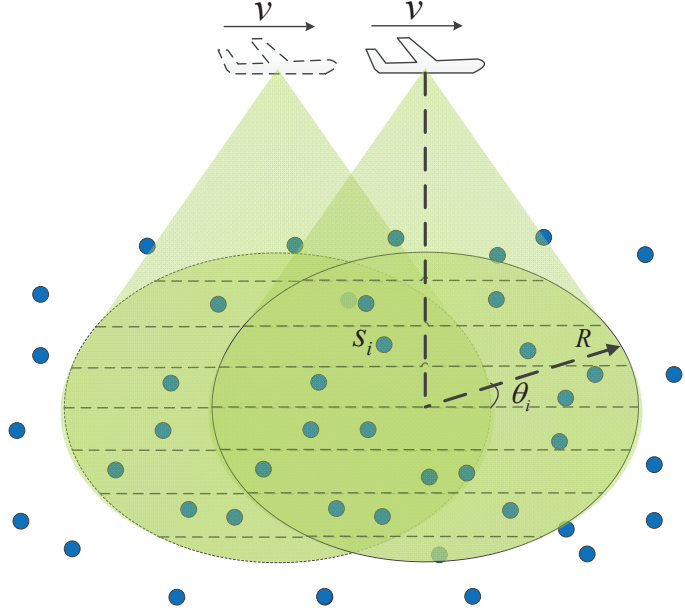}
	\caption{System Model: the UAV flys in a straight trajectory to collects the data from multiple IoT devices on the ground.}
	\label{system model}
\end{figure}

During the data collection process, the UAV broadcasts beacon signal periodically to wake up and synchronize the attached IoT devices on the ground. Once the devices are waked up and covered by the UAV, they will try to access to the UAV followed the CSMA/CA protocol. In addition, for energy efficiency, the UAV is deployed with directional antennas for communications, resulting a circular coverage on the ground, as shown in Fig.~\ref{system model}. Compared with the traditional data collection system with terrestrial infrastructures, the BS deployed on the UAV moves continuously and its coverage changes along with the UAV's trajectory. Therefore, the devices can not connect with the UAV for a long continuous time. Moreover, the UAV-Devices communication durations are also different under the straight trajectory because of devices' different locations. Specifically, for device $s_i$, the communication duration can be expressed as:
\begin{equation}
{T_i} = \frac{{2R\cos {\theta _i}}}{v},
\end{equation}
where $R$ is the radius of the UAV coverage, $v$ is the velocity of the UAV, and ${\theta _i} \in (0,\pi /2)$ is the locations of the device and can be calculated as ${\theta _i} = \arcsin ({y_i}/R)$.

Obviously, the devices located right below the flight trajectory (where $\theta=0$) can connect with the UAV for the longest durations.

\subsection{ Classic Markov Chain Model}
For easy understanding of our cluster division and quitting probability, a classic Markov chain model in traditional CSMA/CA protocol with static BS is presented briefly.

Generally, the CSMA/CA access scheme employs a binary exponential backoff technique, which can be regarded as a Markov chain. For the retry limit $J$, there are $J+1$ backoff stages (from 0 to $J$) in total. The $j$th ($j \in (0,1,...,L)$) stage includes $2^jW_0$ states, where $W_0$ is the initial contention window size in the CSMA/CA protocol. In the $j$th backoff stage , the backoff time is uniformly chosen in the range of $(0,W_{j}-1)$, where $W_{j}=2^jW_0$.

At the beginning of each transmission, the packet will be in the first backoff stage and begins to backoff. During the backoff process, the value of the backoff counter will be chosen from $(0,W_0-1)$. Here we define $q$ as the probability that a device senses the channel busy in a backoff stage. The bakoff counter decreases if the channel is sensed idle and stops when the channel is busy. Accordingly, the value of the backoff counter decreases by one with a probability $(1-q)$ or keeps the same value with a probability $q$ at each time slot. When the backoff counter is reduced to 0, the packet will be transmitted successfully if the channel is sensed idle. Otherwise, the packet transfers to the next backoff stage and repeats the process described above. The contention window size is doubled in the next backoff stage up to its maximum $2^JW_0$. For example, the contention window size is $2^jW_0$ in the $j$th stage and it will become $2^{j+1}W_0$ in the next stage.

When the packet is in the last backoff stage, if the backoff time counter decreases to 0, the packet will be transmitted successfully with an idle channel. Otherwise, this packet will be discarded instead of entering the next stage because the stage $J$ is the last one.

\subsection{The Division of Device Clusters}\label{cluster}
As mentioned before, the devices in different locations can access the UAV for different time durations. The access process following CSMA/CA is modeled as a Markov process. The times of each device traversing all the states in the Markov process is critical for its transmission. The more times a devices traversing the whole Markov chain, the higher probability of successful transmission.

Therefore, we divide the devices into different band-shaped clusters paralleled to the flight trajectory according to the times that they can traverse the whole Markov process, as shown in Fig.\ref{system model}. One device will be allocated into the cluster $\mathbb{C}_i$ if it can traverse the Markov process for $m_i$ times. Here $m_i$ can be expressed as
\begin{equation}
m_i = \left\lfloor \frac{{T_i}}{\Delta} \right\rfloor = \left\lfloor \frac{{2Rcos\theta_i}}{v\times\Delta} \right\rfloor,
\end{equation}
where $\left\lfloor x \right\rfloor$ represents the rounding down to $x$. $\Delta$ is the time cost for a packet to traverse all the Markov stages. According to \cite{xiao}, $\Delta$ is shown as
\begin{equation}\label{deta}
\begin{split}
\Delta  = &E({B})\delta  + E({F})[\frac{{{P_{s}}}}{{{P_{b}}}}{T_s} + \frac{{({P_{b}} - {P_{s}})}}{{{P_{b}}}}{T_c}] \\&+ {J}({T_c} + {T_o}),
\end{split}
\end{equation}
where ${P_{s}}$ and ${P_{b}}$ are the probabilities that the devices transmit a packet successfully, and the probability that there is at least one device transmitting the packet, respectively. ${B}$ represents the total number of backoff counter, and $E({B})$ is the average number of backoff counter during the backoff stage, which can be shown as
\begin{equation}
E({B}) = \sum\limits_{j = 0}^{{J}} {\frac{{{W_{j}} - 1}}{2}}.
\end{equation}

${F}$ is the overall time when the counter freezes, and $E({F})$ is shown as
\begin{equation}
E({F}) = \frac{{E({B})}}{{1 - {q}}}{q}.
\end{equation}

In \eqref{deta}, $T_{s}$ and $T_{c}$  denote the average time that the channel is sensed as busy and the average time for a collision. $T_{o}$ is the time that a device has to wait after the access collides before the next channel sensing.

There are two handshaking techniques, i.e. the two-way handshaking technique (basic access mechanism) and the four-way handshaking technique (RTS/CTS mechanism) in CSMA/CA. Please refer to \cite{b6} for the detailed process of these two mechanisms and we are not allowed to explain it because of the space limitation.
The definitions of $T_s$, $T_c$ and $T_o$ are different in the basic access mechanism and RTS/CTS mechanism. For clarity, we use $T_s^{b}$, $T_c^{b}$, $T_o^{b}$ and $T_s^{r}$, $T_c^{r}$, $T_o^{r}$ to denote the parameters in basic access mechanism and RTS/CTS mechanism, respectively. Then, we have
\begin{equation}
T_s^{b}=T_{\rm{H}}+T_{\rm{E}}+\rm{SIFS}+T_{\rm{ACK}}+\rm{DIFS}+2\times\delta,
\end{equation}
\begin{equation}
T_c^{b}=T_{\rm{H}}+T_{\rm{E}}+\rm{DIFS}+\delta,
\end{equation}
\begin{equation}
T_o^{b}=\rm{SIFS}+T_{\rm{ACK\_timeout}},
\end{equation}
\begin{equation}
\begin{split}
T_s^{r}=&T_{\rm{RTS}}+T_{\rm{CTS}}+T_{\rm{H}}+T_{\rm{E}}+3\times\rm{SIFS}+\\& T_{\rm{ACK}}+\rm{DIFS},
\end{split}
\end{equation}
\begin{equation}
T_c^{r}=T_{\rm{RTS}}+\rm{SIFS}+T_{\rm{ACK}}+\rm{DIFS},
\end{equation}
\begin{equation}
T_o^{r}=\rm{SIFS}+T_{\rm{CTS\_timeout}},
\end{equation}
where $T_{\rm{H}}$, $T_{\rm{E}}$, $T_{\rm{ACK}}$, SIFS, DIFS, $T_{\rm{RTS}}$ and $T_{\rm{CTS}}$ denote the time to transmit the header (including MAC header, physical header), the time to transmit a payload with length E, the time to transmit an ACK, the time durations of a SIFS and DIFS, and the time to transmit a RTS and CTS , respectively. $T_{\rm{CTS\_timeout}}$ and $T_{\rm{ACK\_timeout}}$ denote the duration of the ACK and CTS timeouts respectively. $\delta$ denotes the duration for an idle time slot.

Under this division strategy, the devices located in the center band of the UAV's coverage can connect with the UAV for the longest time. If these devices repeat the backoff process for $N$ times, they belong to the cluster $\mathbb{C}_N$ and $N=\left\lfloor \frac{t_N}{v\times\Delta}\right\rfloor$.
In this way, all the devices in the coverage can be divided into $N$ clusters and each devices in the cluster $\mathbb{C}_i, (i=1,2,...,N)$ can repeat the Markov process for $i$ times.

On the other hand, we assume that the devices are uniformly distributed in the UAV's coverage with density $\rho$, the number of the devices in the area follows the Poission distribution according to \cite{b16}. The probability that there are $n$ devices in the cluster $\mathbb{C}_i$ can be shown as
\begin{equation}\label{eq8}
{f_i}(n) = \frac{{{{(\rho {A_i})}^n}{e^{ - \rho {A_i}}}}}{{n!}},n = 0,1,2,...,\infty,
\end{equation}
where ${A_i}$ is the area of $\mathbb{C}_i$. This distribution can help us to to analysis the saturation throughput of the network in the following section.

\subsection{Quitting Probability}\label{quitting}
Due to UAV's mobility, the devices on the ground will quit the access process when they are out of the coverage. To model this special scenario, a new concept called quitting probability $Q$ is proposed to indicates the probability that any device is out of the UAV's coverage at each time slot. According to the cluster division, we also define $Q_i$ as the quitting probability of devices in cluster $\mathbb{C}_i$.

Since the devices have different communication durations with the UAV, their packets will go through the whole Markov process for different times. The longer they stay in the UAV's coverage, the higher probability they can transmit packets successfully to the UAV because they have more opportunities to repeat the network access. Once a device quits the UAV's coverage, the packet transmitted from it cannot traverse all the Markov states. 

To calculate the quitting probability of the devices in cluster $\mathbb{C}_i$, we first consider a situation in which a packet sent by a certain device traverses all the backoff stages. In this case, the transmission state transfers to state $(J,0)$. Here we use $P_b$ to denote the stationary probability of state $(J,0)$. If the communication link breaks, the packet can not contend for transmission and the Markov state transition process will stop. As a result, the packet can never get to the final state $(J,0)$ and this scenario is regarded as the packet quitting the access process. Under this situation, the probability can be shown as $(1-P_b)$. Therefore, we defined $(1-P_b)$ as the probability that a packet quits the contention process for one time.

For devices in cluster $\mathbb{C}_i$, they can repeat the whole Markov process for $m_i$ times. If a device fails to transmit a packet for all $m_i$ times with a probability $Q_i$, it means that the device can not upload information to the UAV.
Thus, $Q_i$ reflects the probability that a device quits the UAV's coverage at each time slot during the access process, and it can be calculated as
\begin{equation}
{Q_i} = {(1 - {P_b})^{{m_i}}}.
\end{equation}

By introducing the quitting probability into the Markov chain model, we can provide insight to the transmission process in the UAV-based network and build a modified model which is more appropriate for the UAV data collection system.

\newcounter{tmepEq}
\setcounter{tmepEq}{\value{equation}}
\setcounter{equation}{18}
\begin{figure*}[!b]
	\hrulefill
	\begin{equation}\label{long_1}
	{q_i} = 1 - \prod\limits_{h = 1}^{i - 1} {(\sum\limits_{n = 0}^\infty  {{f_h}(n)} {{(1 - {\tau _h})}^{{n_h}}})(\sum\limits_{n = 1}^\infty  {{f_i}(n)} {{(1 - {\tau _i})}^{n - 1}}} )\prod\limits_{h = i + 1}^N {(\sum\limits_{n = 0}^\infty  {{f_h}(n)} {{(1 - {\tau _h})}^{{n_h}}})}
	\end{equation}\vspace{-0mm}
	\begin{equation}\label{long_2}
	{b_{i,j,k}} = \frac{{{W_{i,j}} - k}}{{{W_{i,j}}}}\left\{ {\begin{array}{*{20}{l}}
		{(1 - {q_i})(1 - {Q_i})\sum\limits_{j = 0}^{{J} - 1} {{b_{i,j,0}} + {b_{i,{J},0}},~j = 0} }\\
		{[{q_i}(1 - {Q_i}){\rm{ + }}{Q_i}]{b_{i,j - 1,0}},~j \in [1,{J}]}
		\end{array}} \right.
	\end{equation}
\end{figure*}

\setcounter{equation}{\value{tmepEq}}

\setcounter{equation}{21}
\begin{figure*}[!b]
	\begin{small}
		\begin{equation}\label{long_3}
		{b_{i,0,0}} = \frac{{2{P_a}{P_b}}}{{2{W_{i,0}}{P_{eq}}[{P_a}(1 - {{(2{P_{eq}})}^{{J}}}) - (1 - P_{eq}^{{J}}){P_b}] + 2{P_b}(1 - P_{eq}^{{J} + 1}) + {P_b}[(1 - {q_i})(1 - {Q_i})(1 - P_{eq}^{{J}}) + {P_a}P_{eq}^{{J}}]({W_{i,0}} - 1)}}
		\end{equation}
	\end{small}
\end{figure*}
\setcounter{equation}{\value{tmepEq}}

\section{Modified Markov Chain Model}\label{section:3}
To analyze the saturated throughput of the CSMA/CA in UAV-based air-ground network, we propose a modified Markov chain model to explore the access process of the devices in the network. We divide all the devices in the coverage into $N$ band-shaped clusters along the flight trajectory. In each cluster $\mathbb{C}_i$ ($i=1,...,N$), there are $n_i$ devices, each of which is assumed to have a packet available for transmission. Meanwhile, we assume that the packet collides with constant and independent probability regardless of the number of retransmission suffered.

\begin{figure}[!htp]
	\centering
	\includegraphics [width=8.5cm]{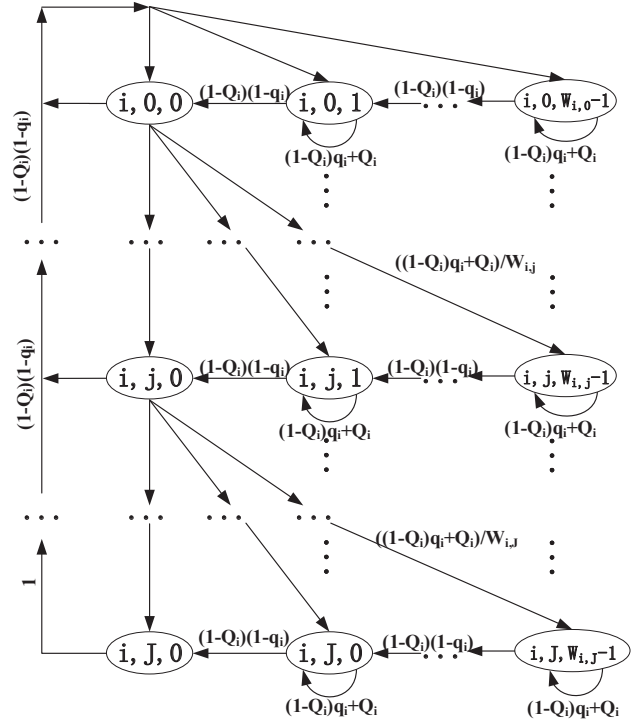}
	\caption{Markov chain model for devices access process in a UAV-based data collection system.}
	\label{Markovchainmodel}
	\vspace{-5mm}
\end{figure}

Compared to the model for the conventional network, we propose a three-dimensional Markov chain model to describe the access process of a packet in UAV-based air-ground network. To clearly describe Markov chain model, we introduce three coefficients ($i$, $j$ and $b(t)$) to represent the Markov process. Here $b(t)$ represents the backoff counter for a certain device at time $t$ and $i$ denotes the number of cluster. Let $s(t)$ be the stochastic process indicating the devices' backoff stage $(j=0,...,J,s(t)=j)$ at time $t$. The backoff stage changes after any unsuccessful access transmission until it reaches the maximum value $J$. $J$ also indicates the last backoff stage and the retry limit. In addition, we define $W_{i,j}$ as the contention window in the $j$th backoff stage for the devices in the cluster $i$. For the first access round, the $W_{i,0}$ equals to $C{W_{\min }}$. After each failed access, contention window will be doubled, up to the maximum $CW_{\rm{max}}$. Therefore, the size of contention window in stage $j$ will be ${W_{i,j}}{\rm{ = }}{2^j}{W_{i,0}},~0 \le j \le {J}$.

Moreover, the quitting probability is adopted in this model. In the classic model, the backoff time counter decreases if the channel is sensed idle. However, in this modified model, the backoff counter decreases if the channel is sensed idle and the device does not quit the UAV's coverage at each time slot. Therefore, the probability that the backoff counter decreases is $(1-Q_i)(1-q_i)$, where $q_i$ is the probability that a device in cluster $\mathbb{C}_i$ senses the channel channel as busy.

The detail stage transition process is shown in Fig.\ref{Markovchainmodel}.
In Fig.~\ref{Markovchainmodel}, the ellipses with three parameters ($i$, $j$ and $b(t)$) represent the states of this Markov process. The arrows show the direction of the state transmission. The formulas along the arrows, like $(1-Q_i)(1-q_i), (1-Q_i)q_i+Q_i$, are the probabilities for one-step transmission. According to Fig.~\ref{Markovchainmodel}, the one-step transmission probabilities of the Markov process is shown below.

\begin{subequations}\label{eq1_1}
	\begin{align}
	&  \{\begin{array}{l}
	P_{i,j,k|i,j,k + 1} =(1-Q_i)(1-q_i)\\
	j \in [0,J],~k \in [0,W_{i,j} - 2]
	\end{array}, \\
	& \{\begin{array}{l}
	{P_{i,j,k|i,j,k}} = (1 - {Q_i}){q_i} + {Q_i}\\
	~j \in [0,{J}],~k \in [1,{W_{i,j}} - 1]
	\end{array},\\
	& \{\begin{array}{l}
	{P_{i,0,k|i,j,0}} = \frac{{(1 - {Q_i})(1 - {q_i})}}{{{W_{i,0}}}}\\
	~j \in [0,{J} - 1],~k \in [1,{W_{i,0}} - 1]
	\end{array},\\
	& \{\begin{array}{l}
	{P_{i,0,k|i,{J},0}} = \frac{1}{{{W_{i,0}}}}\\
	~k \in [1,{W_{i,0}} - 1]
	\end{array},\\
	& \{\begin{array}{l}
	{P_{i,j,k|i,j - 1,0}} = \frac{{(1 - {Q_i}){q_i} + {Q_i}}}{{{W_{i,j}}}}\\
	~j \in [1,{J}],~k \in [0,{W_{i,j}} - 1]
	\end{array},
	\end{align}
\end{subequations}
where $P_{i_1,j_1,k_1|i_0,j_0,k_0}$ is a short notation and it is defined as
\begin{small}
	\begin{equation}
	\begin{split}
	&P_{i,j_1,k_1|i,j_0,k_0}~=\\&~P\{i,~s(t+1)=j_1,~b(t+1)=k_1~|~i,~s(t)=j_0,~b(t)=k_0\}.
	\end{split}
	\end{equation}
\end{small}

The equation (\ref{eq1_1}a) means that the backoff counter decreases in the beginning of each time slot with the probability that the stages of the device stays in the coverage area and the channel is idle at the same time. The equation (\ref{eq1_1}b) tells us the probability that the device stays in the same stages because it quits the coverage or the channel is sensed as busy at each time slot. The equation (\ref{eq1_1}c) and (\ref{eq1_1}d) represent the probability that the backoff stages return to 0 from the former stages. For devices in stages $(i,j,0)$, they will transmit to $(i,0,k)$ after a successful transmission or quit the network. For the stages in $(i,J,0)$, they will return to stages 0. The last equation shows the probability of rescheduling a contention state after an unsuccessful transmission.

\section{Performance Analysis}\label{section:4}

\setcounter{equation}{28}
\begin{figure*}[!b]
	\hrulefill
	\begin{small}
		\begin{equation}\label{eq22}
		{b_{i,0,0}} = \frac{{2{P_a}{P_b}}}{{2{CW_{\rm{min}}^i}{P_{eq}}[{P_a}(1 - {{(2{P_{eq}})}^{{J}}}) - (1 - P_{eq}^{{J}}){P_b}] + 2{P_b}(1 - P_{eq}^{{J} + 1}) + {P_b}[(1 - {q_i})(1 - {Q_i})(1 - P_{eq}^{{J}}) + {P_a}P_{eq}^{{J}}]({CW_{\rm{min}}^i} - 1)}}
		\end{equation}
	\end{small}
\end{figure*}
\setcounter{equation}{15}

This section analyzes the Markov chain process and gives the solution of the saturation throughput.

\subsection{Markov Chain Process Analysis} \label{Markov}
The 3-D Markov process is convergent and the proof is presented in Appendix A. Before the analysis, we first define the stationary distribution of the chain ${b_{i,j,k}}$ as
\begin{equation}
\begin{split}
{b_{i,j,k}} = & \mathop {\lim }\limits_{t \to \infty } P\{ cluster=i,s(t) = j,b(t) = k\},\\& ~j \in [0,J],~k \in [0,{W_{i,j}}].
\end{split}
\end{equation}

Then, for a device in the cluster $i$, we get
\begin{equation}\label{eq6}
{b_{i,j,0}} = {P_{eq}}{b_{i,j - 1,0}} = P_{eq}^j{b_{i,0,0}}, ~{\rm{  j}} \in {\rm{[0,}}{{J}}{\rm{]}},
\end{equation}
where $P_{eq}$ is expressed as
\begin{equation}\label{eq6-1}
P_{eq} = (1 - {Q_i}){q_i} + {Q_i},
\end{equation}
where $q_i$ is the probability that the channel is busy when the device is in the backoff stage.

During the access process, according to the CSMA/CA protocol, once a device in the cluster $\mathbb{C}_i$ connects to the UAV, other devices either in the cluster $\mathbb{C}_i$ or not cannot transmit information to the UAV. Combining the Poission distribution of the number of the devices in \eqref{eq8}, we can calculate the $q_i$ as \eqref{long_1}.


According to the chain regularities, each value of the backoff counter $k \in [1,{W_{i,j}} - 1]$ can be obtained by the equation \eqref{long_2}.

By substituting \eqref{eq6} in \eqref{long_2}, all the values ${b_{i,j,k}}$ can be expressed as a function of ${b_{i,0,0}}$. Then, considering the normalization condition, we can easily obtain:
\setcounter{equation}{20}
\begin{equation}\label{eq12}
\begin{split}
1&= \sum\limits_{j = 0}^{{J}}\sum\limits_{k = 0}^{{W_{i,j}} - 1} {{b_{i,j,k}}} \\&= \sum\limits_{j = 0}^{{J}} {{b_{i,j,0}}}  + \sum\limits_{k = 1}^{{W_{i,j}} - 1} {{b_{i,0,k}}}  + \sum\limits_{j = 1}^{{J}} \sum\limits_{k = 1}^{{W_{i,j}} - 1} {{b_{i,j,k}}},
\end{split}
\end{equation}
where ${P_a} = (1 - {P_{eq}}), {P_b} = (1 - 2{P_{eq}})$. Therefore, we can obtain the expression of $b_{i,0,0}$, as shown in \eqref{long_3}.

According to the Markov chain in Fig. 2, the devices stop the transmission when their backoff counters decrease to 0. Thus, the probability $\tau_i$ that one device in the cluster $\mathbb{C}_i$ transmits a packet in a randomly selected time slot can be shown as:
\setcounter{equation}{22}
\begin{equation}\label{eq111}
{\tau _i} = \sum\limits_{j = 0}^{{L}} {{b_{i,j,0}} = \frac{{1 - P_{eq}^{{J} + 1}}}{{1 - {P_{eq}}}}{b_{i,0,0}}}.
\end{equation}

By combining \eqref{long_1} and \eqref{eq111}, we can calculate the value of parameters $\tau_i$ numerically.

\subsection{Saturation Throughput Computation}

Let $P_{tr}$ be the probability that there is at least one device doing the transmission for a certain time. The probability can be shown as
\begin{equation}\label{eq17}
{P_{tr}} = 1 - \prod\limits_{h = 1}^N {(\sum\limits_{n = 0}^\infty  {{f_h}(n)} {{(1 - {\tau _h})}^n})}.
\end{equation}

For the devices in cluster $\mathbb{C}_i$, the probability of successful transmission $P_{s_i}$ can be calculated as
\begin{equation}\label{eq18}
\begin{split}
{P_{{s_i}}} =& \left[\sum\limits_{n = 1}^\infty  {{f_h}(n)} n{\tau _i}{(1 - {\tau _i})^{n - 1}}\right] \times \\& \prod\limits_{h = 0,h \ne i}^{N - 1} \left[{\sum\limits_{n = 0}^\infty  {{f_h}(n)} {{(1 - {\tau _h})}^n}} \right].
\end{split}
\end{equation}

Therefore, in the whole coverage, the overall probability ${P_s}$ of successful packet transmission to UAV can be shown as ${P_s} = \sum\limits_{i = 1}^{N} {{P_{{s_i}}}}$. Finally, by combining the \eqref{eq17} and \eqref{eq18}, the throughput of the network can be shown as
\begin{equation}\label{eq19}
S = \frac{{{P_s}{P_{tr}}E[P]}}{{(1 - {P_{tr}})
		\sigma  + {P_s}{P_{tr}}{T_s} + {P_{tr}}(1 - {P_s}){T_c}}}.
\end{equation}

\section{Modified CSMA/CA Protocol}\label{section:5}

In this UAV-based data collection network, the UAV moves continuously because of its aerodynamic characteristics and the large area of the IoT devices, causing the unfair access time among devices in different locations.
According to the performance analysis above, we have insight about how the network parameters affect the saturation throughput according to the equations in Section~\ref{section:3}. Therefore, we fully consider the characteristics of this special network and propose a modified CSMA/CA protocol that is expected to provide a more fair access control scheme for the devices.

In our new protocol, we adaptively adjust the size of the contention window ([$CW_{\rm{min}}, CW_{\rm{max}}$]) for each device in different cluster. Under this scheme, for the devices with short communication duration, we set lower contention window size to make them access the channel within shorter time.
Without loss of generality, the contention window size for those which have longer communication duration is larger because they have more time and opportunities to participate in the access process. The larger contention window size make them wait a little longer during each transmission, which gives more opportunities to those devices with shorter communication time to access the channel.

Here we modify the values of $CW_{\rm{min}}$ and $CW_{\rm{max}}$ to change the contention window size. The binary exponential backoff mechanism is also adopted in this model. According to this mechanism, $CW_{\rm{max}}=2^{J}\times CW_{\rm{min}}$. Thus, we only need to change the values of $CW_{\rm{min}}$ and $J$ for each cluster. According to the performance analysis above, the durations of access process for devices in each cluster are limited and distinct. Thus, we adjust the initial contention window size and the retry limit of devices in each cluster according to their communication durations.

Let $CW_{\rm{min}}^{\rm{max}}$ be the maximum value of $CW_{\rm{min}}$ for devices in cluster $\mathbb{C}_1$ and $T$ be the longest communications duration for  $\mathbb{C}_1$. Then, the initial contention window size for devices in cluster $\mathbb{C}_i$ is calculated as:
\begin{equation}\label{eq20}
CW_{\rm{min}}^i=\left\lceil (1-\frac{t_i}{T})\times CW_{\rm{min}}^{\rm{max}} \right\rceil,
\end{equation}
where $t_i=i\times\Delta$ is the communications duration for cluster $\mathbb{C}_i$. $\left\lceil x \right\rceil$ denotes the upper integer which is the closest to $x$.

On the other hand, we hope that the devices in cluster $\mathbb{C}_N$ have the largest contention window size. Thus, we assign a retry limit $J_{\rm{{max}}}$ in cluster $N$. Similarly, the values of the retry limit in other clusters should be smaller than $J_{\rm{{max}}}$. Therefore, for the cluster $\mathbb{C}_i$, the value of retry limit is expressed as
\begin{equation}\label{eq21}
J_i=\left\lceil J_{\rm{{max}}}\times\frac{t_i}{T} \right\rceil.
\end{equation}

Finally, the saturation throughput of this modified protocol can be calculated using our analytical model proposed above. Under this situation, $b_{i,0,0}$ is recalculated as \eqref{eq22}.
In addition, $\tau_i$ is redefined as:
\setcounter{equation}{29}
\begin{equation}\label{eq23}
{\tau _i} = \sum\limits_{j = 0}^{{J_i}} {{b_{i,j,0}} = \frac{{1 - P_{eq}^{{J_i} + 1}}}{{1 - {P_{eq}}}}{b_{i,0,0}}}.
\end{equation}

By combining \eqref{long_1} and \eqref{eq23}, the value of $\tau_i$ can be obtained numerically. Then, according to \eqref{eq17}, \eqref{eq18} and \eqref{eq19}, we can get the saturation throughput of this modified protocol.

\section{Simulation and Numerical Results} \label{section:6}

In this section, we validate our system model firstly by comparing the theoretical results with those obtained from the Monte Carlo simulation. Then, the effects of some network parameters, such as UAV's velocity and the density of the devices, on the throughput are analyzed. An idle time slot $\delta $ is 50$\rm{\mu} s$, ${\rm{ACK}_{\rm{timeout}}}$ = 300 $\rm{\mu} s$, ${\rm{CTS}_{\rm{timeout}}}$ = 300 $\rm{\mu} s$, ACK = 112 $\rm{\mu} s$, RTS = 160 $\rm{\mu} s$, CTS = 112 $\rm{\mu} s$, SIFS = 28 $\rm{\mu} s$, DIFS = 128 $\rm{\mu} s$. Assume the channel data rate is equal to $1 \rm{Mbit/s}$ and the frame size , denoted by EP, is fixed at 8k Bytes. The coverage radius is assumed to be 1000 m unless stated otherwise.
\begin{figure}[t]
	\centering
	\includegraphics [width=9cm]{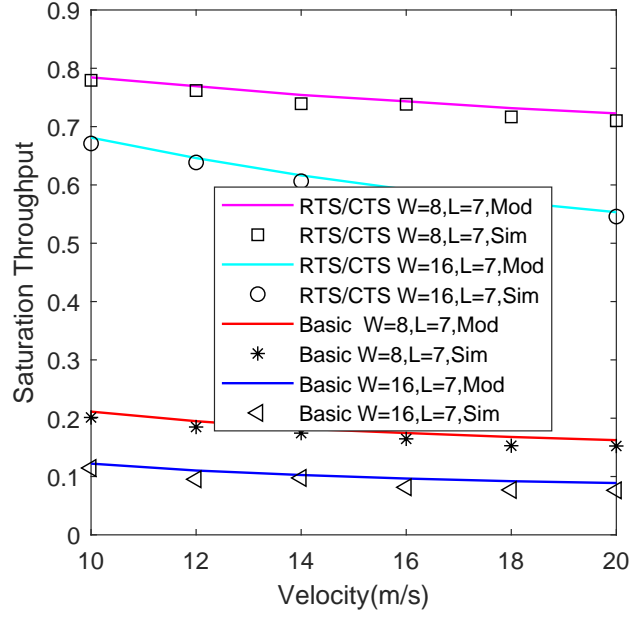}
	\caption{Saturation throughput vs. the velocity of the UAV for CSMA/CA protocol.}
	\label{validation}
\end{figure}

Fig.~\ref{validation} plots the theoretical and simulation saturation throughput as a function of UAV's velocity under different parameters, marked as `Mod' and `Sim' in legends respectively.
To fully verify the network performance, we validate the results in both basic access and RTS/CTS cases.
The user density in this simulation is $\rho = 50/\rm{{km}^2}$ and the retry limit is $J = 7$ for all devices.
In addition, we adopt two values of the initial contention window size ($CW_{\rm{{min}}} = 8,16$) to explore the network performance.
It is clear to see that the theoretical results (lines) match the simulation results (symbols) well, which perfectly validates our model.

Moreover, the saturation throughput of proposed model decreases along with the increase of the UAV's velocity. This trend is expected, since the devices connect with the UAV for shorter time when the UAV flies faster.
In this situation, it is hard for all the devices to access the channel in short communication durations, resulting in the lower throughput. From Fig.~\ref{validation} we can also get some interesting observations that RTS/CTS access scheme always has better performance than the basic access scheme.
This is because, compared with the basis mechanism, RTS/CTS is designed to combat the problem of hidden terminal and it can reduce the duration of a collision to improve the system performance.
Therefore, RTS/CTS is more likely to perform better than the basic access mechanism, especially when long messages are transmitted.
\begin{figure}[t]
	\centering
	\includegraphics [width=9cm]{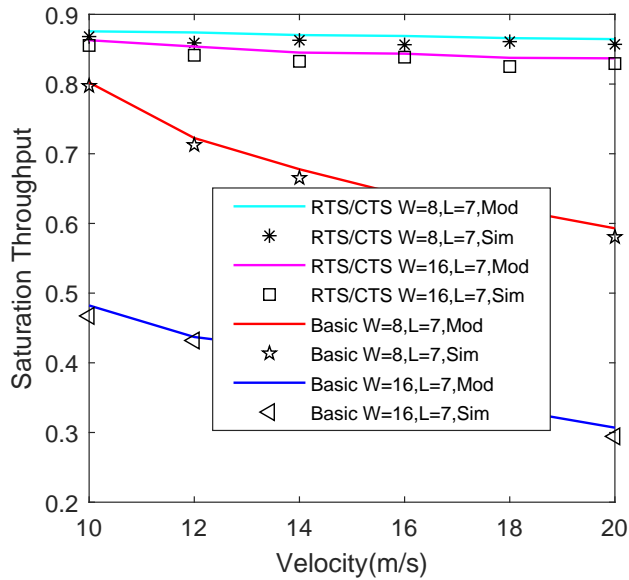}
	\caption{Saturation throughput vs. the velocity of the UAV for modified CSMA/CA protocol.}
	\label{validation2}
\end{figure}

In order to validate the theoretical results of our modified CSMA/CA, we also implement some experiments using the same parameters adopted in Fig.~\ref{validation}.
The theoretical and simulation results of the modified protocol are presented in Fig.~\ref{validation2}. Similarly, both the basic access mechanism and RTS/CTS are considered.
As expected, the curves of our theoretical results also fit the simulation results very well. It means that the theoretical model can also be applied to analyze the throughput performance of the modified protocol.
In addition, comparing the results in Fig.~\ref{validation} and Fig.~\ref{validation2}, it can be concluded that the modified protocol improves the throughput enormously.
Specifically, for the basic access mechanism, the maximal throughput is a little larger than 0.2 when $v=10 \rm{m/s}$, as shown in Fig.~\ref{validation}.
However, the modified protocol improves throughput to 0.8.
Meanwhile, the maximal throughput for RTS/CTS mechanism in Fig.~\ref{validation} is no more than 0.8, but the throughput in Fig.~\ref{validation2} is improved to above 0.85 after adopting the modified protocol.
\begin{figure}[t]
	\centering
	\includegraphics [width=9cm]{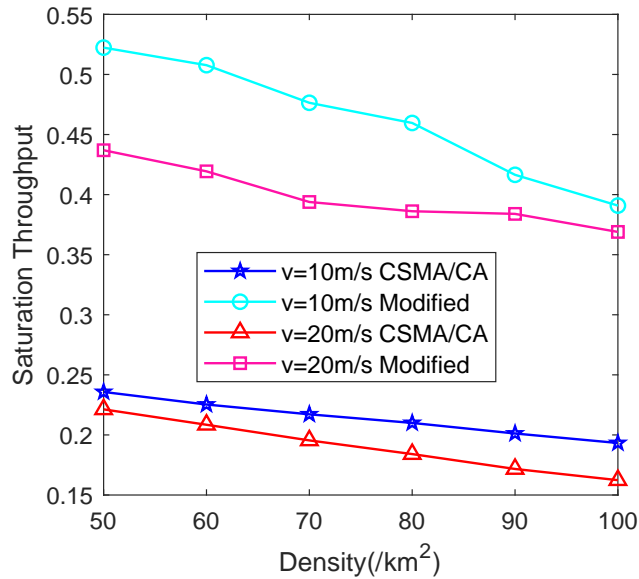}
	\caption{Saturation throughput vs. the density of the devices for basic access mechanism ($CW_{\rm{{min}}} = 8, J = 7, \rm{EP} = 8k~\rm{Bytes}$, $R$ = 1000 m).}
	\label{den_basic}
\end{figure}

In Fig.~\ref{den_basic}, we plot the normalized saturation throughput in terms of the density of the devices for basic access mechanism under two different UAV's velocities ($v = 10, 20m/s$).
Results for both conventional CSMA/CA and the modified protocol are presented.
From Fig.~\ref{den_basic}, we can see that, as the devices' density increases, the throughput decreases correspondingly.
For the spectrum resource limitation, the more devices participate in the access process simultaneously, the more access collision will happen, leading to the lower throughput.

In addition, comparing the throughput of those two protocol, we see that the modified protocol performs better than the conventional CSMA/CA in throughput, since the new protocol provides a fair contending environment which alleviates the congestion among devices in the UAV's coverage.
\begin{figure}[t]
	\centering
	\includegraphics [width=9cm]{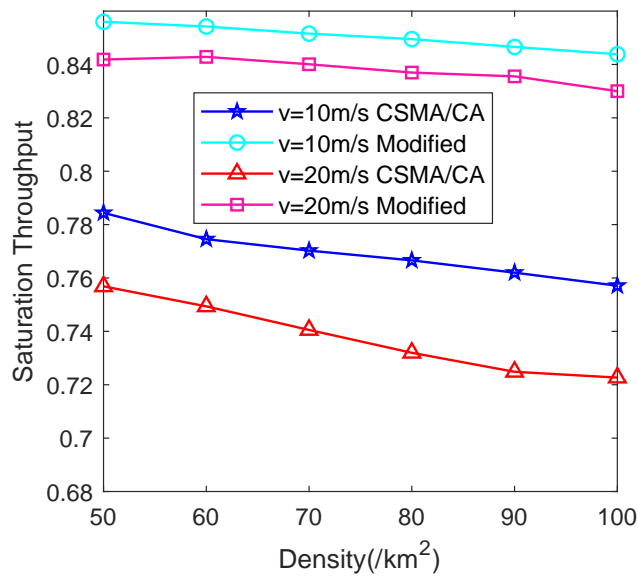}
	\caption{Saturation throughput vs. the density of the devices for RTS/CTS mechanism 
	($CW_{\rm{{min}}} = 8, J = 7, \rm{EP} = 8k~\rm{Bytes}$, $R$ = 1000 m).}
	\label{dens}
\end{figure}

Fig.~\ref{dens} shows the saturation throughput of the system for RTS/CTS mechanism as a function of the density of the devices when $v$ = 10 m/s and 20 m/s. In this case, device's density has similar impacts on throughput performance compared with the results in Fig.~\ref{den_basic}.
Additionally, the modified protocol can also improve the throughput performance for the RTS/CTS mechanism.
In detail, the improvement of the throughput is near to 0.1 compared with basic mechanism when $\rho = 80/\rm{km^2}$.
In addition, the modified MAC protocol is more robust to density of the devices for the RTS/CTS mechanism. The difference of the throughput from $50/\rm{km^2}$ to 100/$\rm{km^2}$ is less than 0.01 under the modified protocol.
The throughput only changes 0.01 along with the increase of device's density (from $50$ to $100/\rm{km^2}$).
\begin{figure}[t]
	\centering
	\includegraphics [width=9cm]{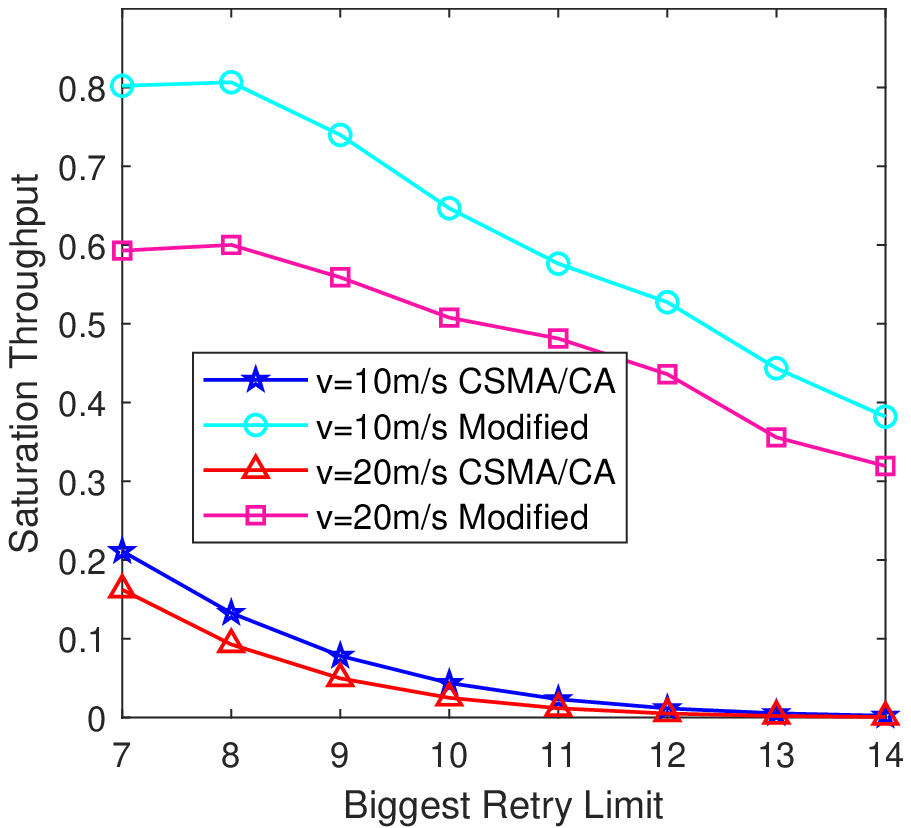}
	\caption{Throughput for different retry limit for basic access mechanism ($CW_{\rm{{min}}} = 8, \rm{EP} = 8k~\rm{Bytes}$, $R$ = 1000~m, $\rho = 50/\rm{km^2}$).}
	\label{re_basic}
\end{figure}

\begin{figure}[t]
	\centering
	\includegraphics [width=9cm]{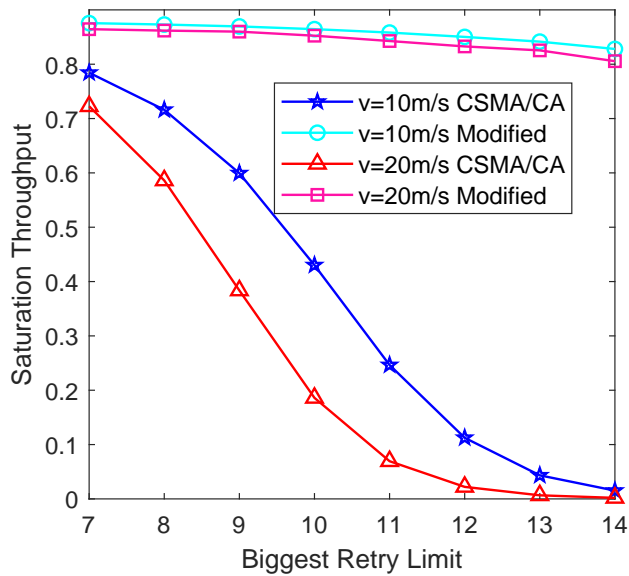}
	\caption{Throughput for different retry limit for RTS/CTS mechanism ($CW_{\rm{{min}}} = 8, \rm{EP} = 8k~\rm{Bytes}$, $R$ = 1000~m, $\rho= 50/\rm{km^2}$).}
	\label{re}
\end{figure}

Moreover, we also implement some experiments to analyze the influence of the retry limit on the saturation throughput for both basic access and RTS/CTS mechanisms.
The biggest retry limit on the horizontal axis in Fig.~\ref{re_basic} and Fig.~\ref{re} is the retry limit for the devices in cluster $\mathbb{C}_N$.
We assume that the retry limit for all the devices in the traditional CSMA/CA protocol is the same and equal to the biggest one.
While in the modified protocol, the retry limit for devices in different clusters is different.

We observe that the increase of the retry limit (from 7-14) has negative impacts on the throughput.
The reason is that the devices have more opportunities to retry the transmission with larger retry limit, causing more collision and resulting in lower throughput.
Furthermore, the modified protocol is reliable when the biggest retry limit increases for both basic access and RTS/CTS mechanism.
Specially, it has better performance in RTS/CTS mechanism. The throughput in basic access mechanism declines to the half of the maximum when the retry limit increases to 14.
However, the throughput just falls a little in RTS/CTS mechanism when the retry limit increases to 14.

\begin{figure}[t]
	\centering
	\includegraphics [width=9cm]{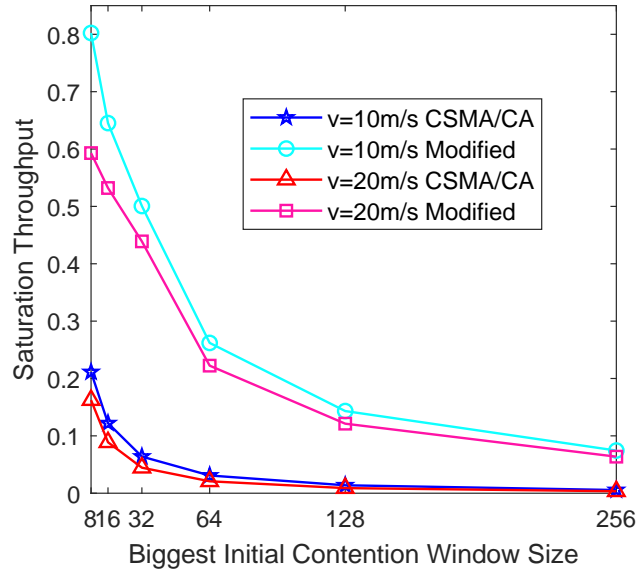}
	\caption{Throughput for different initial contention window size for basic access mechanism ($J = 8, \rm{EP} = 8k~\rm{Bytes}$, $R$ = 1000~m, $\rho= 50/\rm{km^2}$).}
	\label{cw_basic}
\end{figure}

\begin{figure}[t]
	\centering
	\includegraphics [width=9cm]{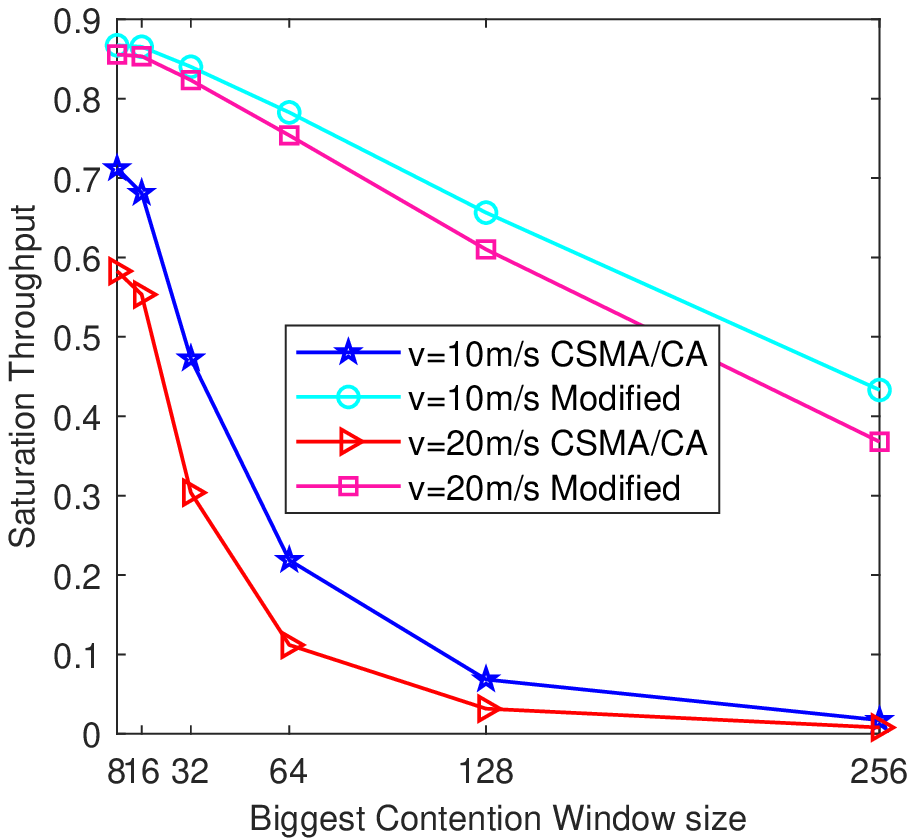}
	\caption{Throughput for different initial contention window size for RTS/CTS mechanism ($J = 8, \rm{EP} = 8k~\rm{Bytes}$, $R$ = 1000~m, $\rho= 50/\rm{km^2}$).}
	\label{cw}
\end{figure}

The impacts of the initial contention window sizes are analyzed for both basic access mechanism and RTS/CTS mechanism, as shown in Fig.~\ref{cw_basic} and Fig.~\ref{cw}.
For conventional CSMA/CA protocol, the initial contention windows size for all the devices are the same.
However, they are different in the modified protocol according to the equations in Section~\ref{section:5}.
The biggest initial contention window size is set for the devices in the cluster $\mathbb{C}_1$ and that for other devices is obtained from \eqref{eq20}.
The horizontal axis of Fig.~\ref{cw_basic} and Fig.~\ref{cw} denotes the biggest initial contention window size.
Comparing the results in Fig.~\ref{cw_basic} and Fig.~\ref{cw}, we see that the initial contention window size affects the throughput seriously.
For the basic access mechanism, the throughput declines to almost zero when the initial contention window size climbs to 256. For the RTS/CTS mechanism, the effects cannot be ignored either. The throughput drops to below 0.1 in this scenario.
\begin{figure}[t]
	\centering
	\includegraphics[width=9cm]{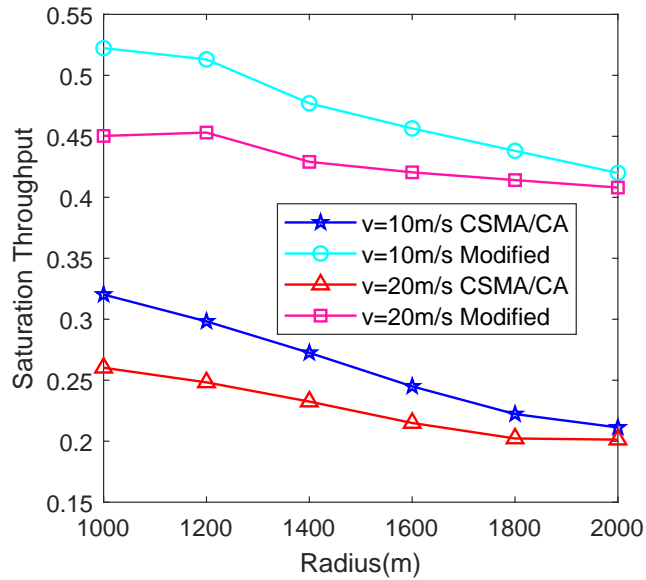}
	\caption{Throughput in terms of different coverage radius for basic access mechanism($J = 8, \rm{EP} = 8k~\rm{Bytes}$, $CW_{\rm{{min}}} = 8$, $\rho = 50/{km}^2$).}
	\label{r_basic}
\end{figure}

\begin{figure}[t]
	\centering
	\includegraphics[width=9cm]{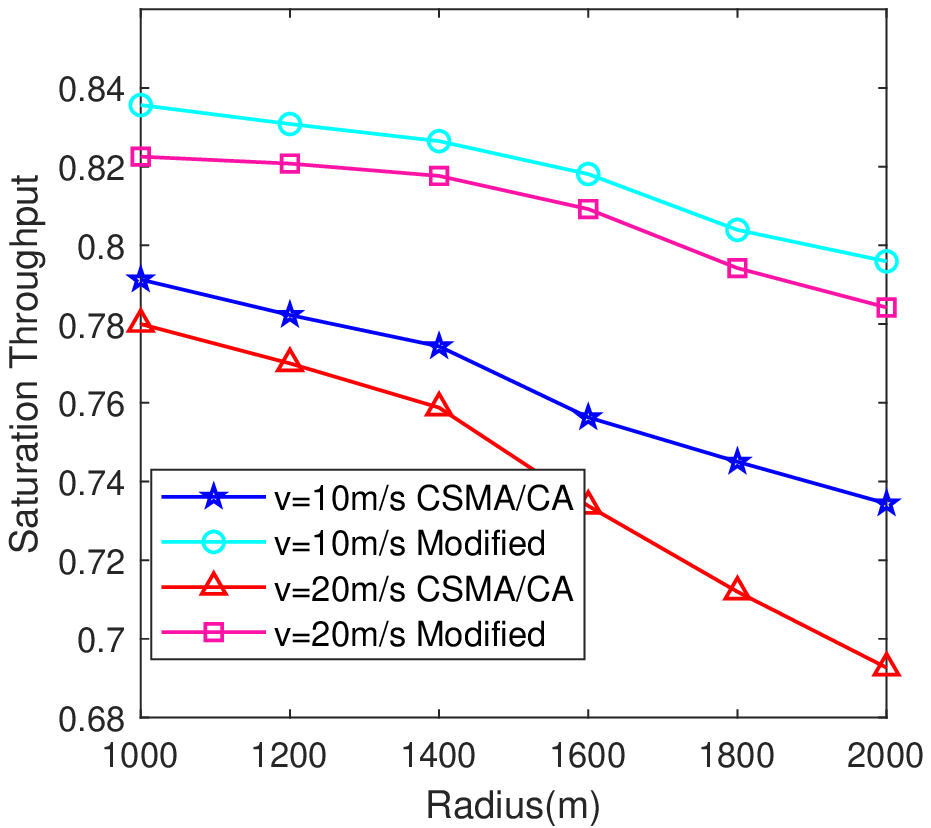}
	\caption{Throughput in term of different coverage radius for RTS/CTS mechanism ($J = 8,\rm{EP} = 8k~\rm{Bytes}$, $CW_{\rm{{min}}} = 8$, $\rho = 50/{km}^2$).}
	\label{r}
\end{figure}
The reason is that if the initial contention window increases, the devices will spend more time in the backoff process. Therefore, the channel is more likely to stay in idle state, leading to the waste of the channel resource.
On the other hand, the transmission delay will be longer with larger contention window size which also influences the throughput performance.
On the contrary, the modified protocol still gives better throughput though the initial contention window size increases to 256. In this case, the throughput for basic access mechanism and RTS/CTS is around 0.1 and 0.4 respectively, which are much better than those of the conventional protocol.

Finally, the impacts of UAV's coverage are considered. Since the UAV's coverage is a circle, only the effects of the radius of the coverage need to be considered.
Here we set the density of the devices as $\rho = 50/{Km}^2$, and the radius of the coverage increases from 1000 m to 2000 m. The results for basic access mechanism and RTS/CTS mechanism are given in Fig.~\ref{r_basic} and Fig.~\ref{r} respectively.

The throughput falls off along with the increase of the radius for both the basic access mechanism and the RTS/CTS mechanism.
Larger coverage area means that more devices are able to be covered by UAV and transmit the packet simultaneously, resulting in more congestion and the falloff of the system throughput.
Though both increasing the density of the devices and UAV's coverage radius enlarge the number of the devices that the UAV serves, there is still some difference between these two methods.
Increasing the device's density means only the number of the devices increases.
However, larger coverage not only increases the number of the devices, but also improves the communication durations between the devices and the UAV.
Thus, the results in Fig.~\ref{den_basic} and Fig.~\ref{r_basic}, Fig.~\ref{dens} and Fig.~\ref{r} are different.
As expected, the modified protocol is also reliable when coverage radius increases and gives better performance than the conventional one.

\section{Conclusion}\label{section:7}
In our paper, a scenario where a UAV flies straightly to collect data from a set of IoT devices is considered.
Based on the characteristics of the UAV-based network, the devices in UAV's coverage is divided into different clusters and a quitting probability is proposed. The proposed theoretical model is proved accurate for the performance analysis of CSMA/CA protocol in UAV-based system.
Then, a modified protocol based on CSMA/CA is proposed that fully considers the heterogeneity of the access time to adaptively allocate the time resource among the devices within the difference clusters. This modified protocol enormously improves the system performance.
We also analyze the impacts of different network parameters including retry limit, initial contention window size, UAV's velocity, the density of the devices, and the UAV's coverage.

\begin{appendices}\label{app_a}
\section{Proof of convergence of the Markov chain model}
The convergence of a Markov process indicates that there is a stationary distribution of the Markov process. Here we prove that the modified Markov process proposed in our paper is convergent by illustrating that there is only one stationary distribution for this process. 

For clear presentation, we first give some definitions about Markov chain. 
\newtheorem{definition}{Definition}
\begin{definition}\label{definition1}
	Let $\{X_n,n\geq0\}$ be a discrete-time stochastic process, whose state space is denoted by $S=\{1,2...\}$. If for any $n\geq0$, and $i_0,...,i_n,i_{n+1} \in S$, we have
	\begin{equation}
	\begin{split}
	P(X_{n+1}=i_{n+1}|X_0=i_0, X_1=i_1,...,X_n=i_n) \\ =P(X_{n+1}=i_{n+1}|X_n=i_n)
	\end{split}
	\end{equation}
\end{definition}
where $n$ denotes the number of the time slot.

\newtheorem{definition2}{Definition}
\begin{definition}\label{definition2}
	(1)Let $\{X_n,n\geq0\}$ be a discrete-time Markov chain with the state space $S$, then for any $n\geq0$,
	\begin{equation}\label{eqa1}
	p_{ij}(n,n+1)=P(X_{n+1}=j|X_n=i), i,j\in S
	\end{equation}
	where $p_{ij}$ is the one-step transition probability of the Markov chain.
	
	(2) If the transition probability $p_{ij}(n,n+1), i,j \in S$, defined in \ref{eqa1}, are independent of n, then it will be denoted by $p_{ij}^{(1)}$ and the corresponding $\{X_n, n\geq0\}$ is called a homogeneous Markov chain.
\end{definition}
According to the definitions above, the Markov chain proposed in our paper is a homogeneous Markov chain with finite states.

Next, we give the definition of period of a state.
\newtheorem{definition3}{Definition}
\begin{definition}\label{definition3}
	Let $\{X_n,n \geq 0\}$ be a Markov chain with state space $S$. For any state $j \in S$, if $\{n:p_{jj}^{(n)}\}\not=\emptyset$
	\begin{equation}\label{eq3}
	d(j)=GCD\{n:p_{jj}^{n}\textgreater0\}
	\end{equation}
\end{definition}

where $GCD$ is the \textit{great common divisor}, $p_{jj}^n$ denotes the probability state $j$ returns to itself after $n$ steps.
$d(j)$ is called the period of state $j$. Moreover, state $j$ is called aperiodic if $d(j)=1$.

According to this definition, the states in our Markov process are all aperiodic and the Markov chain is aperiodic.

Next, we give some lemmas and the detailed proofs of those lemmas can be found in \cite{book1},\cite{book2}.
\newtheorem{lemma}{Lemma}
\begin{lemma}\label{lemma1}
	A Markov chain is called irreducible if its state space $S$ is a communicating class. If the chain has finite states, the communicating class is recurrent.
\end{lemma}
\newtheorem{lemma2}{Lemma}
\begin{lemma}\label{lemma2}
	For a finite aperiodic irreducible Markov chain model, it has a stationary distribution.
\end{lemma}
\newtheorem{lemma3}{Lemma}
\begin{lemma}\label{lemma3}
	A Markov chain has a unique stationary distribution if and only if there is only one positive-recurrent communicating class.
\end{lemma}

Assume the state space $S_1$ of our proposed Markov process. For all the states $i,j\in S_1$, $i$ can transit to $j$ and $j$ can get to state $i$. It means $S_1$ is a communicating class. Therefore, according to Lemma \ref{lemma1} and all the definitions listed, the Markov chain is a aperiodic irreducible Markov chain with finite states and there is only one positive recurrent communicating class. Then, according to Lemma \ref{lemma2} and Lemma \ref{lemma3}, we can conclude that there is only one stationary distribution for the proposed Markov process. Thus, the convergence of the Markov process is proved.
The stationary distribution is denoted by \eqref{eq6} and \eqref{long_2}. 

\end{appendices}
\bibliographystyle{ieeetran}
\bibliography{tvt}
\begin{IEEEbiography}[{\includegraphics[width=1in,height=1.25in,clip,keepaspectratio]{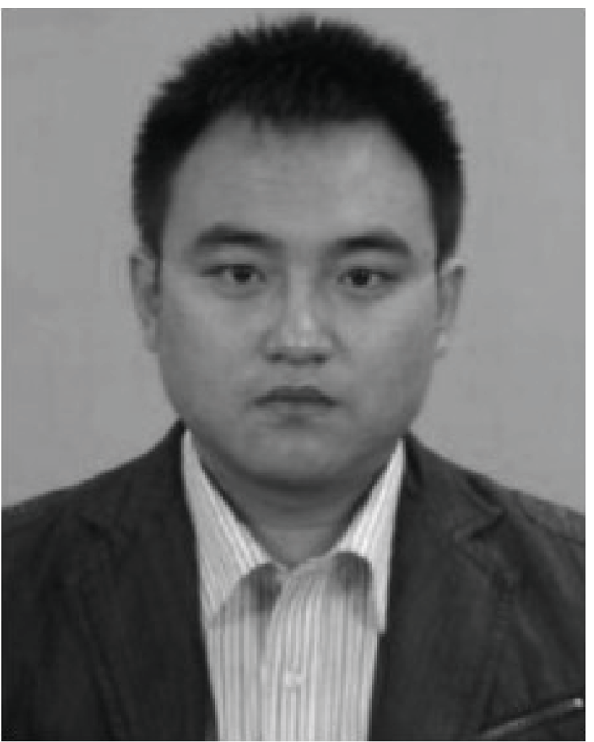}}]{Bin Li}
	(S’12–M’14) received the B.S. and Ph.D. degrees from Xi’an Jiaotong University, Xi’an, China, in 2006 and 2014, respectively, both in electrical and electronics engineering. Since 2014, he has been with the Department of Communication Engineering, Northwestern Polytechnical University, Xi’an, China, and he is currently an Associate Professor. His current research interests include the Internet of Things, network coding, MIMO, and wireless channel measurement modeling.
\end{IEEEbiography}
\begin{IEEEbiography}[{\includegraphics[width=1in,height=1.25in,clip,keepaspectratio]{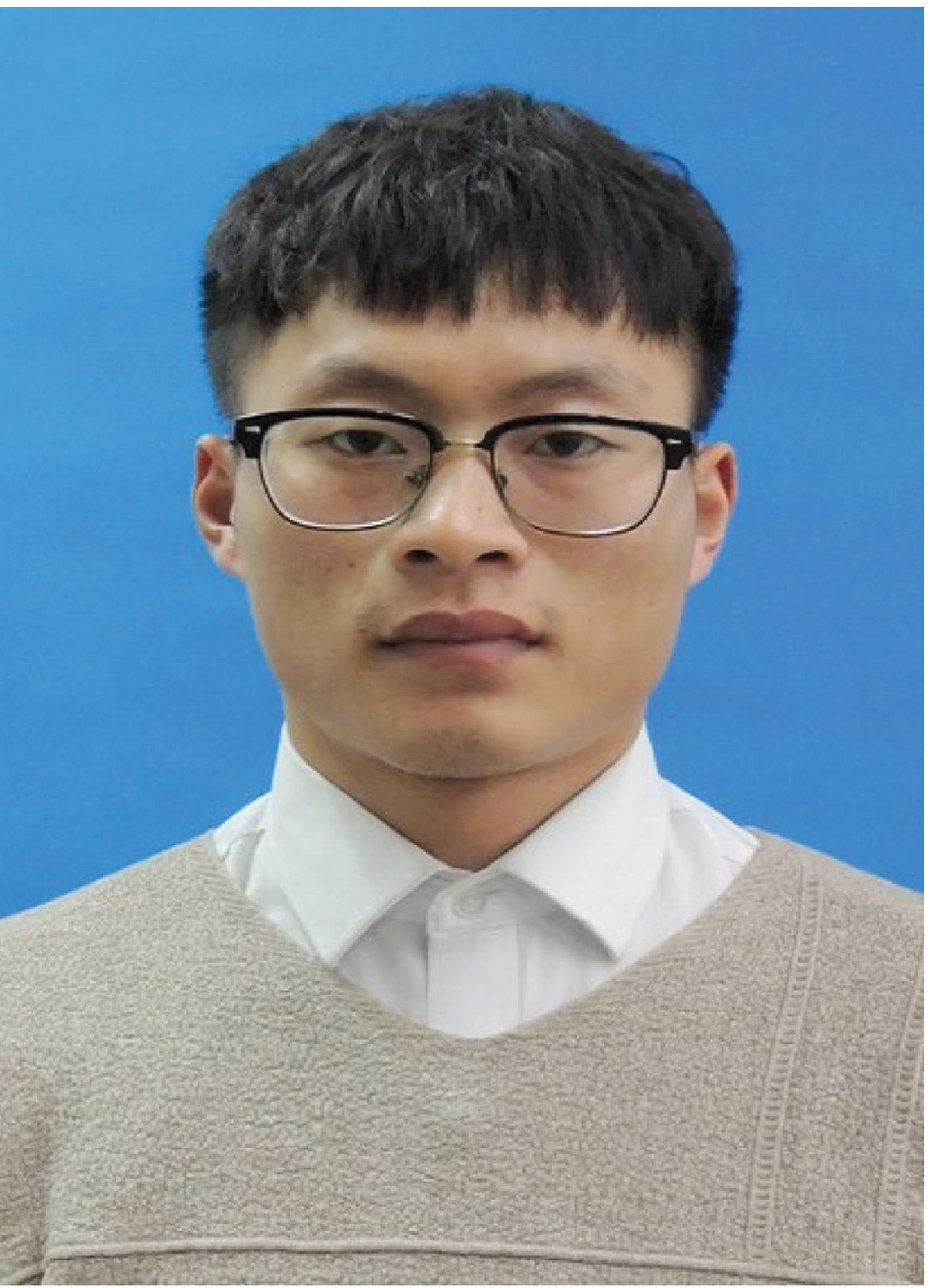}}]{Xianzhen Guo}
	 received the B.E. degree in communication engineering form Northwestern Polytechnical University, Xian, China, in 2019, where he is currently working toward the master's degree with the Department of Communication Engineering. His research focuses on UAV communications, convex optimization.
\end{IEEEbiography}
\begin{IEEEbiography}[{\includegraphics[width=1in,height=1.25in,clip,keepaspectratio]{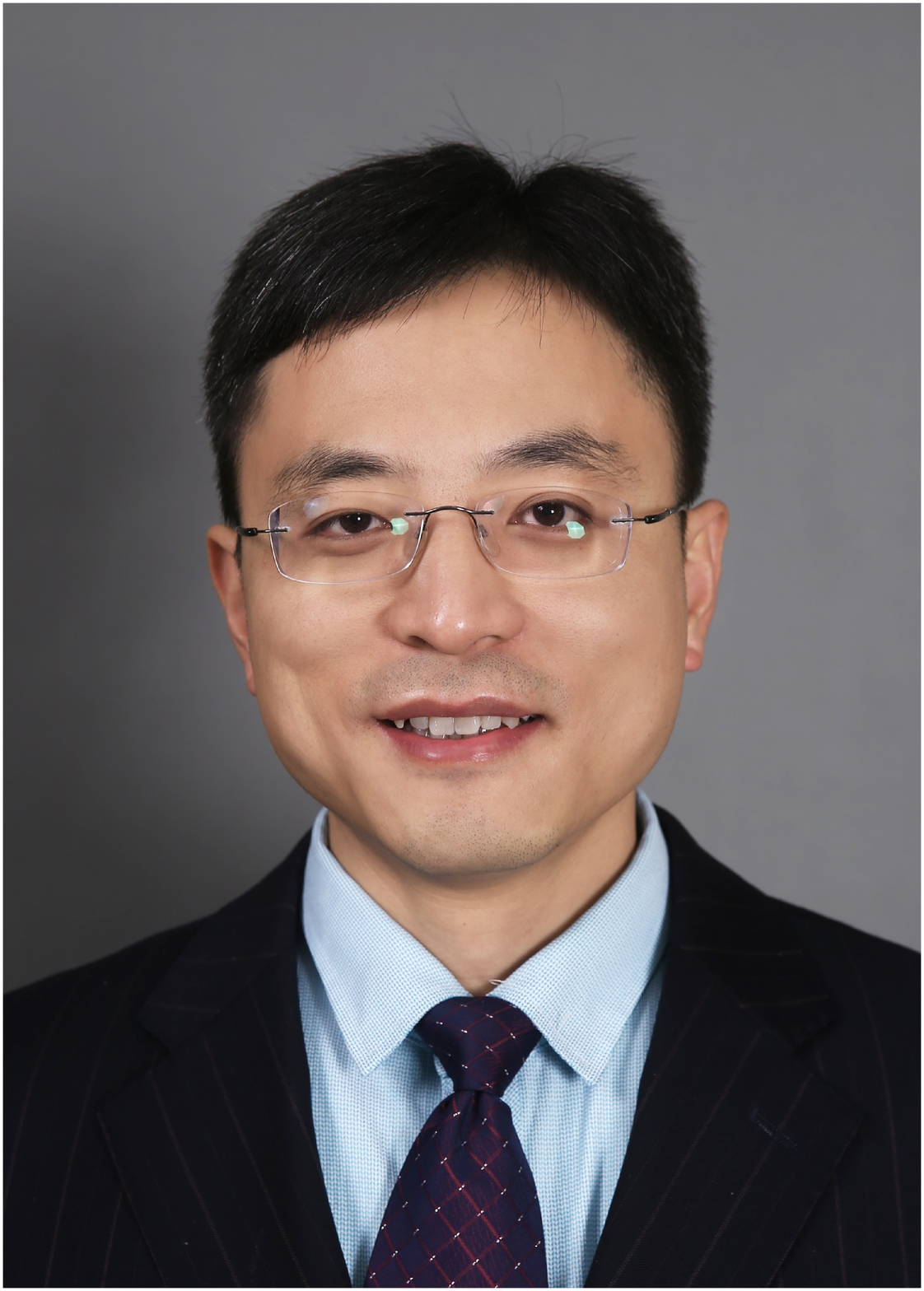}}]{Ruonan Zhang}
	 (S’09–M’10) received the B.S. and M.Sc. degrees in electrical and electronics engineering from Xian Jiaotong University, Xian, China, in 2000 and 2003, respectively, and the Ph.D. degree in electrical and computer engineering from the University of Victoria, Victoria, BC, Canada, in 2010.,He was an IC Architecture Engineer with Motorola Inc., Chicago, IL, USA, and Freescale Semiconductor Inc., Tianjin, China, from 2003 to 2006. Since 2010, he has been with the Department of Communication Engineering, Northwestern Polytechnical University, Xian, China, where he is currently a Professor. His current research interests include wireless channel measurement and modeling, architecture and protocol design of wireless networks, and satellite communications.,Dr. Zhang was a recipient of the New Century Excellent Talent Grant from the Ministry of Education of China and the Best Paper Award of IEEE NaNA 2016. He has served as a Local Arrangement Co-Chair for the IEEE/CIC International Conference on Communications in China in 2013 and as an Associate Editor for the Journal of Communications and Networks.
\end{IEEEbiography}
\begin{IEEEbiography}[{\includegraphics[width=1in,height=1.25in,clip,keepaspectratio]{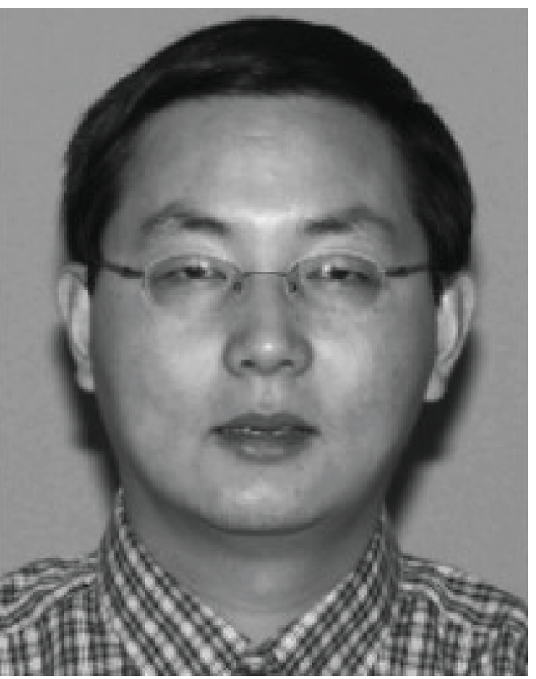}}]{Xiaojiang Du}
	(Fellow, IEEE) received the B.S. and M.S. degrees in electrical engineering from Tsinghua University, Beijing, China, in 1996 and 1998, respectively, and the M.S. and Ph.D. degrees in electrical engineering from the University of Maryland at College Park, in 2002 and 2003, respectively. He is currently a tenured Professor with the Department of Computer and Information Sciences, Temple University, Philadelphia, USA. He has authored more than 300 journal and conference papers in these areas and a book published by Springer. He has been awarded more than 6 million U.S. dollars research grants from the US National Science Foundation (NSF), Army Research Office, Air Force Research Laboratory, NASA, Qatar, the State of Pennsylvania, and Amazon. His research interests are security, wireless networks, and systems. He is a Life Member of ACM. He received the Best Paper Award at IEEE GLOBECOM 2014 and the Best Poster Runner-Up Award at ACM MobiHoc 2014. He serves on the editorial boards of three international journals.
\end{IEEEbiography}
\begin{IEEEbiography}[{\includegraphics[width=1in,height=1.25in]{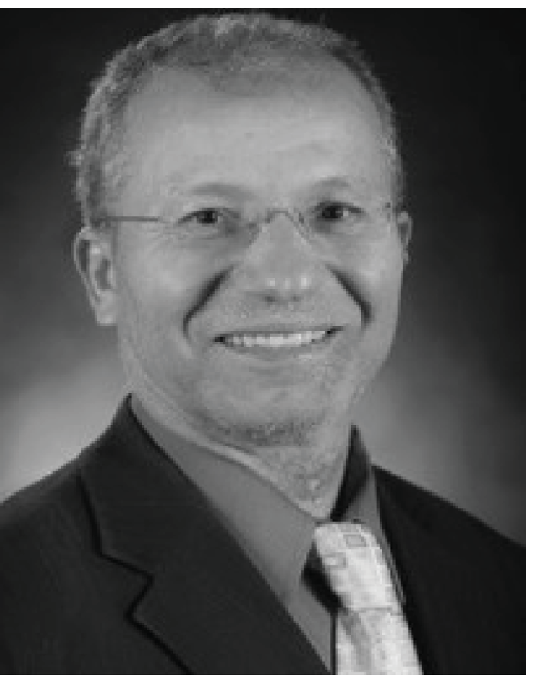}}]{Mohsen Guizani}
	(S’85–M’89–SM’99–F’09) received the B.S. (Hons.) and M.S. degrees in electrical engineering and the M.S. and Ph.D. degrees in computer engineering from Syracuse University, Syracuse, NY, USA, in 1984, 1986, 1987, and 1990, respectively. He served in different academic and administrative positions at the University of Idaho, Western Michigan University, the University of West Florida, the University of Missouri–Kansas City, the University of Colorado at Boulder, and Syracuse University. He is currently a Professor with the CSE Department, Qatar University, Qatar. His research interests include wireless communications and mobile computing, computer networks, mobile cloud computing, security, and smart grid. He has authored 9 books and more than 500 publications in refereed journals and conferences. He is a Senior Member of ACM. He also served as a member, chair, and general chair of a number of international conferences. Throughout his career, he received three teaching awards and four research awards. He also received the 2017 IEEE Communications Society WTC Recognition Award and the 2018 AdHoc Technical Committee Recognition Award for his contribution to outstanding research in wireless communications and Ad-Hoc Sensor networks. He was the Chair of the IEEE Communications Society Wireless Technical Committee and the Chair of the TAOS Technical Committee. He served as the IEEE Computer Society Distinguished Speaker. He is currently the IEEE ComSoc Distinguished Lecturer. He guest edited a number of special issues in IEEE journals and magazines. He is currently the Editor-in-Chief of the IEEE Network Magazine, serves on the editorial boards of several international technical journals, and the Founder and Editor-in-Chief of Wireless Communications and Mobile Computing (Wiley).
\end{IEEEbiography}
\end{document}